\documentclass[aps,prb,twocolumn,showpacs,preprintnumbers,amsmath,amssymb,superscriptaddress]{revtex4}

\usepackage{graphicx}% Include figure files
\usepackage{dcolumn}% Align table columns on decimal point
\usepackage{bm}% bold math
\usepackage{float}

\begin{document}

\title{The influence of spin-flip scattering on the preparation and detection of a single spin state in a quantum dot
attached to a spin battery}

\author{Piotr Trocha}
\email{ptrocha@amu.edu.pl}\affiliation{Department of Physics,
Adam Mickiewicz University, 61-614 Pozna\'n, Poland}

\date{\today}

\begin{abstract}
Recently,
the possibility of an all electrical scheme of preparation and readout for a single spin state in
a single quantum dot attached to
spin biased leads has been shown
[F. Chi \emph{et al} Phys. Rev. B {\bf 81}, 075310 (2010)]. However, spin scattering mechanisms have been omitted. To
remedy this lack we consider the
influence of the spin-flip scattering process on the proposed preparation and readout scheme.
\end{abstract}

 \pacs{72.25.-b, 73.23.-b, 73.63.Kv, 85.35.Be}

\maketitle

\section{Introduction}\label{Sec:1}
Coherent manipulation of a single spin state in quantum dots (QDs) is
crucial for spintronics and quantum information physics.
A qubit, which is the basic requirement for quantum information processing,
can be realized as an electron's spin on a single-electron quantum dot\cite{loss1}.
Quantum dots seem to be promising elements for spin-based quantum computer
as their properties are rather easily and electrically controllable.
Preparation and manipulation of a single spin state in a QD has been realized
by various optical and magnetic means\cite{koppens,press,xu,nowack} or
spin to charge conversion\cite{elzerman,hanson} techniques.
However, to our knowledge,
there is no experiment allowing all-electrical control of a single spin state
in a QD (altough such an attempt should reduce the complexity of the experiment).

Recently, Chi \emph{et al} have proposed a theoretical scheme for the
preparation and readout of a single spin state in a QD utilizing spin battery\cite{chi}
realized in recent experiments\cite{frolov1,frolov2}.
Manipulation and detection of a spin state have been also investigated
in a double quantum dot system coupled to the spin\cite{lu} or charge\cite{fujisawa,taylor} biased
leads.
However, spin scattering mechanisms in a single QD have not been taken into account\cite{chi}.
Intradot spin-flip processes can dramatically change a dot's transport characteristics and
lead to suppression of the tunnel magnetoresistance\cite{rudzinski1,rudzinski2}. At low temperature
the two most important
mechanisms in the QD responsible for spin-flip phenomena are spin-orbit interactions\cite{nazarov} (SOI) and
the hyperfine interaction\cite{johnson} (HI) which
couples the electron's spin to an effective magnetic
field induced by nuclear spins. However, for $B\gg B_{nuclear}\approx 3 mT$
the HI mechanism is suppressed\cite{amasha} and the spin relaxation is then mainly due to
spin-orbit interaction. Many low temperature experiments have been realized for measuring the spin relaxation time
in semiconductor quantum dots\cite{elzerman,hanson,johnson,tarucha,meunier,heiss}.
Although all the ingredients of the QD's spin-based quantum computer are available,
it is still a great challenge to build such a device, accessible for commercial applications.

The electron spin state can relax due to thermal decay,
which prevents spin-based quantum computation
at room temperature. Another dot's spin-flip mechanisms can be induced
by higher order tunneling events. Namely, when the dot is in the Coulomb
blockade regime, co-tunneling processes may flip the spin on a dot\cite{weymann}.
However, incoherent processes, like inelastic transitions and co-tunneling,
provide rather a minor contribution to the spin-flip mechanism\cite{koppens}.
It is also worth mentioning that in the case of strong coupling to the leads
at low temperature
significant impact on the spin scattering mechanism can originate from Kondo resonance.

In this paper we analyze the influence of the intradot spin-flip mechanism on the control
of a single spin state in a quantum dot attached to a spin battery. A spin battery\cite{frolov1,frolov2,hirsch} allows to
inject pure spin current into the dot\cite{wang2}.

The paper is organized as follows. In section \ref{Sec:2} we
describe the model and theoretical formalism. Numerical results
are presented and discussed in section
\ref{Sec:3}. A summary and final conclusions are gathered in section
\ref{Sec:4}.

\section{Model and theoretical formalism}\label{Sec:2}
We consider a single-level quantum dot coupled to one spin battery and one normal lead. Then the Hamiltonian of the system is of the form:
\begin{eqnarray}\label{eq1}
\hat{H}=\sum_{\mathbf k\alpha\sigma} \varepsilon_{{\mathbf
k}\alpha \sigma}c^{\dagger}_{{\mathbf k}\alpha \sigma} c_{{\mathbf
k}\alpha \sigma}+
\sum_{\sigma}\limits\epsilon_{\sigma}q^\dag_{\sigma}q_{\sigma}
+Un_{\sigma}n_{\bar{\sigma}}
       \nonumber\\
+R(q^{\dagger}_{\uparrow}q_{\downarrow}+q^{\dagger}_{\downarrow}q_{\uparrow})
+\sum_{{\mathbf
k}\alpha}\sum_{\sigma}\limits(V_{\sigma}^\alpha
   c^\dag_{{\mathbf k}\alpha\sigma}q_{\sigma}+\rm h.c.).\ \
\end{eqnarray}
The first term describes the left ($\alpha=L$) and right ($\alpha=R$) lead in the
non-interacting quasi-particle approximation. Here,
$c^{\dagger}_{\mathbf{k}\alpha\sigma}$
($c_{\mathbf{k}\alpha\sigma}$) is the creation (annihilation)
operator of an electron with the wavevector $\mathbf{k}$ and spin
$\sigma$ in the lead $\alpha$, whereas
$\varepsilon_{\mathbf{k}\alpha\sigma}$ denotes the corresponding
single-particle energy. The next three terms in the Hamiltonian
(\ref{eq1}) describe the quantum dot. Here,
$n_{\sigma}=q^\dag_{\sigma}q_{\sigma}$ is the particle number
operator ($\sigma=\uparrow ,\downarrow$),
$\epsilon_{\sigma}$ is the discrete energy level of the QD
and $U$ is the intra-dot Coulomb integral.
The parameter $R$ in the Hamiltonian (\ref{eq1})
describes the spin-flip transition amplitude.
%%%%%
The spin-flip term is assumed to be coherent, in the sense that
the spin-flip strength $R$ involves reversible transitions between
up- and down-spin states on the dot. Such an effect may
originate from a spin-orbit coupling\cite{nazarov} in the dot or from the
transverse component of a local magnetic field applied, for
example, within the electron spin resonance technique\cite{loss2}.
%%%%%%
The last term of Hamiltonian (\ref{eq1}) describes
electron tunneling between the leads and dot, where
$V_{\sigma}^\alpha$ are the relevant tunneling matrix elements.
Coupling of the dots to external leads can be parameterized in
terms of $\Gamma^\alpha_{\sigma}(\epsilon)=2\pi|V_{\sigma}^\alpha|^2\rho_{\alpha}$,
where $\rho_{\alpha}$ is the density of states in the $\alpha$th lead.
In the wide band approximation $\Gamma^\alpha_{\sigma}$ is constant within
the electron band,
$\Gamma^\alpha_{\sigma}(\epsilon)=\Gamma^\alpha_{\sigma}={\rm
const}$ for $\epsilon\in\langle-W/2,W/2\rangle$, and
$\Gamma^\alpha_{\sigma}(\epsilon)=0$ otherwise. Here, $W$ denotes
the electron bandwidth.

To find equations governing the dynamics of the system in the
weak coupling regime we follow Ref.\cite{dong} except as regards the derivation of the dot's Green's functions. In contrast to Ref.\cite{dong} we do not restrict our model to weak spin-flip strengths by calculating the dot's Green's functions assuming vanishing $R$. Here, the dot's Green's functions are obtained for finite $R$.
  Details of the method are displayed in the
Appendix. For the sake of simplicity we assume spin degenerate dot energy level
($\epsilon_{\uparrow}=\epsilon_{\downarrow}\equiv\epsilon_{0}$).
The rate equations for dot's expectations values of the density matrix elements acquire
the following form:
\begin{widetext}
\begin{eqnarray}\label{eq2}
 \dot{\rho}_{00}&=&\frac{1}{2}\sum_{\sigma}\{[F_{\sigma}^{-}(\epsilon_0+R)+F_{\sigma}^{-}(\epsilon_0-R)]\rho_{\sigma\sigma}-
[F_{\sigma}^{+}(\epsilon_0+R)+F_{\sigma}^{+}(\epsilon_0-R)]\rho_{00}
+[F_{\sigma}^{-}(\epsilon_0+R)-F_{\sigma}^{-}(\epsilon_0-R)]\rho_{\bar{\sigma}\sigma}\}
 \nonumber \\
\dot{\rho}_{\sigma\sigma}&=&\frac{1}{2}\{[F_{\sigma}^{+}(\epsilon_0+R)+F_{\sigma}^{+}(\epsilon_0-R)]\rho_{00}
-\left[F_{\sigma}^{-}(\epsilon_0+R)+F_{\sigma}^{-}(\epsilon_0-R)+F_{\bar{\sigma}}^{+}(\epsilon_0+U+R)
+F_{\bar{\sigma}}^{+}(\epsilon_0+U-R)\right]\rho_{\sigma\sigma}
\nonumber \\
&+&[F_{\bar{\sigma}}^{-}(\epsilon_0+U+R)+F_{\bar{\sigma}}^{-}(\epsilon_0+U-R)]\rho_{22}
+2iR\rho_{\sigma\bar{\sigma}}
\nonumber \\
&-&[F_{\sigma}^{-}(\epsilon_0+R)-F_{\sigma}^{-}(\epsilon_0-R)-F_{\bar{\sigma}}^{+}(\epsilon_0+U+R)+
F_{\bar{\sigma}}^{+}(\epsilon_0+U-R)+2iR]\rho_{\bar{\sigma}\sigma}\},
\nonumber \\
\dot{\rho}_{\sigma\bar{\sigma}}&=&\frac{1}{4}\{[F^{+}(\epsilon_0+R)-F^{+}(\epsilon_0-R)]\rho_{00}
-[F^{-}(\epsilon_0+R)-F^{-}(\epsilon_0-R)+4iR]\rho_{\bar{\sigma}\bar{\sigma}}
\nonumber \\
&+&[F^{+}(\epsilon_0+U+R)-F^{+}(\epsilon_0+U+R)+4iR]\rho_{\sigma\sigma}
\nonumber \\
&-&[F^{-}(\epsilon_0+R)+F^{-}(\epsilon_0-R)+
F^{+}(\epsilon_0+U+R)+F^{+}(\epsilon_0+U-R)]\rho_{\sigma\bar{\sigma}}\}.
\nonumber \\
\dot{\rho}_{22}&=&\frac{1}{2}\sum_{\sigma}\{[F_{\sigma}^{+}(\epsilon_0+U+R)+F_{\sigma}^{+}(\epsilon_0+U-R)]
\rho_{\bar{\sigma}\bar{\sigma}}
%\nonumber \\
-[F_{\sigma}^{+}(\epsilon_0+U+R)-F_{\sigma}^{+}(\epsilon_0+U-R)]\rho_{\sigma\bar{\sigma}}
\nonumber \\
&-&[F_{\sigma}^{-}(\epsilon_0+U+R)+F_{\sigma}^{-}(\epsilon_0+U-R)]\rho_{22}\}.
\end{eqnarray}
\end{widetext}
In above equations, $F^{\pm}=F_{\uparrow}^{\pm}+F_{\downarrow}^{\pm}$ with
$F_{\sigma}^{+}(x)=\sum_{\alpha}\Gamma_{\alpha\sigma}f(x-\mu_{\alpha\sigma})$
and
$F_{\sigma}^{-}(x)=\sum_{\alpha}\Gamma_{\alpha\sigma}[1-f(x-\mu_{\alpha\sigma})]$,
for
$x=\{\epsilon_\sigma\pm R,\epsilon_\sigma+U\pm R\}$. Here, $f(\epsilon-\mu_{\alpha\sigma})$ is
the Fermi-Dirac distribution function in the $\alpha$th lead.
The thermally averaged diagonal elements of the density matrix
($\rho_{00},\rho_{\sigma\sigma},\rho_{22}$)
refer to the probability of the dot's state being empty, singly occupied by
an electron with spin $\sigma$ and doubly occupied, respectively.
In turn, the nondiagonal terms $\rho_{\sigma\bar{\sigma}}$ describe
coherent transitions between $\sigma$ and $\bar{\sigma}$ spin states on the dot.
Rate equations (\ref{eq2}) together with the completeness relation
$\rho_{00}+\sum_\sigma\rho_{\sigma\sigma}+\rho_{22}=1$ give us full
information about the dynamics of the system. The spin-dependent dot's occupation
numbers are expressed in the following way $n_{\sigma}=\rho_{\sigma\sigma}+\rho_{22}$.

Current flowing from $\alpha$ lead to the dot is obtained from the standard definition:
\begin{eqnarray}\label{eq18}
J_\alpha^j=-e\langle\dot{N}_{\alpha}\rangle=-i\frac{e}{\hbar}\langle
[H,N_{\alpha}]\rangle,
\end{eqnarray}
where $N_\alpha$ is a occupation number operator in $\alpha$ lead.
After performing onerous calculations the current formula,
in the weak coupling approximation,
acquires the following form:
\begin{widetext}
\begin{eqnarray}\label{eq20}
J_\alpha&=&\frac{e}{2\hbar}\sum_{\sigma}\limits\Re\{[F_{\alpha\sigma}^{+}(\epsilon_0+R)+F_{\alpha\sigma}^{+}(\epsilon_0-R)]\rho_{00}-
[F_{\alpha\sigma}^{-}(\epsilon_0+R)+F_{\alpha\sigma}^{-}(\epsilon_0-R)]\rho_{\sigma\sigma}
\nonumber \\
&+&[F_{\alpha\sigma}^{+}(\epsilon_0+U+R)+F_{\alpha\sigma}^{+}(\epsilon_0+U-R)]\rho_{\bar{\sigma}\bar{\sigma}}-
[F_{\alpha\sigma}^{-}(\epsilon_0+U+R)+F_{\alpha\sigma}^{-}(\epsilon_0+U-R)]\rho_{22}
\nonumber \\
&-&[F_{\alpha\sigma}^{-}(\epsilon_0+R)-F_{\alpha\sigma}^{-}(\epsilon_0-R)+
F_{\alpha\sigma}^{-}(\epsilon_0+U+R)-F_{\alpha\sigma}^{-}(\epsilon_{0}+U-R)]\rho_{\bar{\sigma}\sigma}\},
\end{eqnarray}
\end{widetext}
where $\Re[A]$ denotes the real part of $A$.
Finally, current passing through the dot can be symmetrized in the
following way: $J=(J_{L}-J_R)/2$.
The spin current is defined in the
following way: $J_s=(\hbar/2e)(J_{\uparrow}-J_{\downarrow})$, where
$J_{\sigma}$ is the spin-$\sigma$ electron contribution to the charge current.
%%%%%%%%%%%%%%%%%%%%%%%%%%%%%%%%%%%%%%%%%%%%%%%%%%%
\section{Results}\label{Sec:3}
In numerical calculations we take the parameters from Ref.\onlinecite{chi}.
As the charging energy is the largest energy in the QD we choose it
as the energy unit, i.e., $U=1$. Then, other parameters of the system are expressed in units of
the Coulomb interaction parameter $U$. We set the dot-lead coupling to be $\Gamma_L=\Gamma_R=0.02$.
Moreover, we assume that only the left lead provides spin bias, whereas the right electrode is
the normal lead. Specifically, we set $\mu_{L\uparrow}=eV_s$, $\mu_{L\downarrow}=-eV_s$ and
$\mu_{R\uparrow}=\mu_{R\downarrow}=\mu_R=0$, where $V_s$ is the applied spin bias.
No (charge) bias voltage is applied.

\begin{figure}[t]
\begin{center}
  \includegraphics[width=0.9\columnwidth]{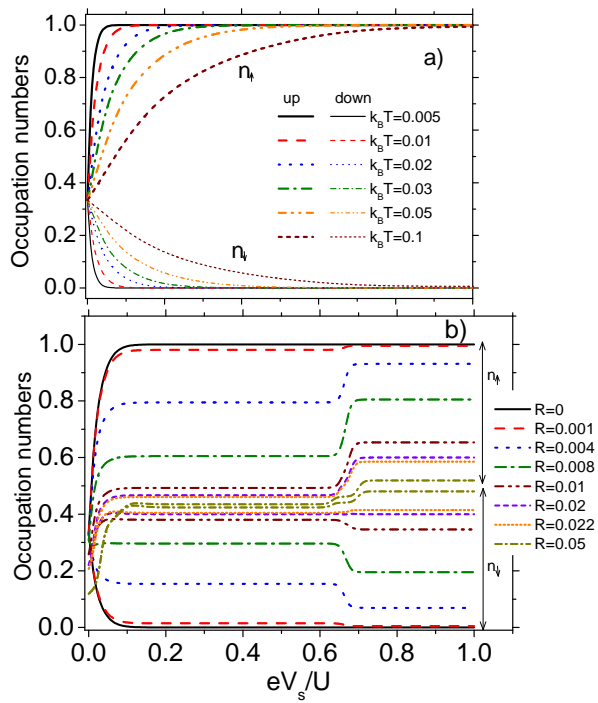}
  \caption{\label{Fig:1}
Dot's occupation numbers as a function of the spin bias
  voltage calculated (a) in the absence of the spin-flip relaxation ($R=0$)
  and for the indicated values of the temperature, (b)
  for the indicated values of parameter $R$ and for temperature $k_BT=0.01$.}
\end{center}
\end{figure}
Following Ref.\cite{chi} we start from analyzing the preparation stage in the stationary limit.
In this case we set the left hand sides of Eqs.(\ref{eq2}) equal to zero, i.e., $\dot{\rho}_{ij}(t)=0$.
Steady state master equations together with the completeness relation
allow us to obtain relevant expectation values of the density matrix elements and
dot occupation numbers of each spin component ($n_{\sigma}$ for $\sigma=\uparrow,\downarrow$).
At the beginning we calculated $n_{\sigma}$ in the absence of spin-flip processes, which is
equivalent to making the assumption that $R=0$. Figure \ref{Fig:1}(a) shows dot's occupation
numbers as a function of the applied spin bias calculated for the dot level situated
symmetrically between $\mu_R$ and $\mu_{L\downarrow}$ ($\epsilon_0=-eV_s/2$).
In such a prepared system QD tends to be occupied by an electron with spin up when a sufficiently
large spin bias is applied. With increasing temperature one needs
to apply greater spin bias to obtain a QD in the state with
a spin-up electron with probability close to unity. This indicates the possibility of
preparation of a single spin state in a quantum dot by using a spin bias
which has been shown in Ref.\cite{chi}. Moreover, a sufficiently
high spin bias is now available in experiments\cite{frolov1,frolov2}.
However, spin-flip processes may have a significant impact on the above picture.
Thus, now we consider the influence of these processes on the preparation stage.
In Fig.\ref{Fig:1}(b) we plot dot occupation numbers for different values of the
parameter $R$ (proportional to the strength of the spin-flip processes).
Generally, when $R$ is nonzero the occupation number $n_{\uparrow}$ ($n_{\downarrow}$)
is reduced (enhanced). However for small $R$ and sufficiently large spin bias
($eV_s\gtrsim 0.68$) this suppression of $n_{\uparrow}$ (enhancement of $n_{\downarrow}$)
can be partially recovered. Thus, if the intradot
spin relaxation is weak enough we are able to prepare dot in
a given state. On the other hand, when the spin-flip processes are strong enough
($R>\Gamma$) the dot can be occupied with an electron with spin up
or down with almost equal probability. Then, even increase in the spin bias voltage
is not able to enforce the dot's occupation by a spin-up electron.
In contrast to the weak spin-flip case, now, further increase
of the spin bias leads to increase of both  $n_{\uparrow}$ and $n_{\downarrow}$.
\begin{figure}[t]
\begin{center}
  \includegraphics[width=0.9\columnwidth]{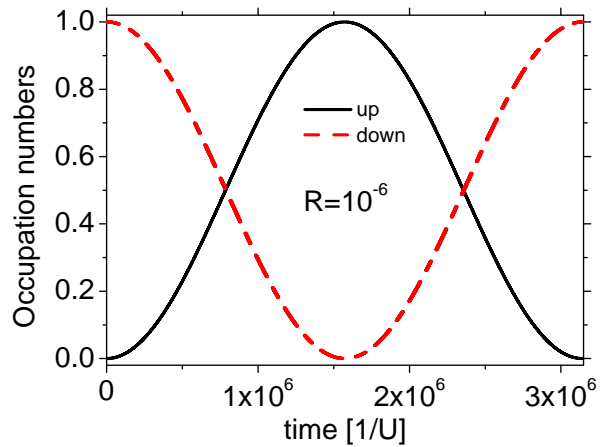}
  \caption{\label{Fig:2}
Time evolution of the dot's occupation numbers
calculated for initial state $|0,1\rangle$
and for the indicated value of parameter $R$, and for $\epsilon_0=-0.5$, $k_BT=0.01$.}
\end{center}
\end{figure}
Moreover, additional features in the dot's occupation number dependence
can be noticed. These features are due to splitting of the dot's energy level
induced by a sufficiently strong spin-flip strength i.e., for $R\gg\Gamma$.
After the initialization stage we lower the dot's energy level
in such way as to prohibit sequential tunneling events.
%%%%%%TABLE
%
%
%\begin{widetext}
\begin{figure*}[t]
\begin{center}
  \includegraphics[width=0.82\textwidth]{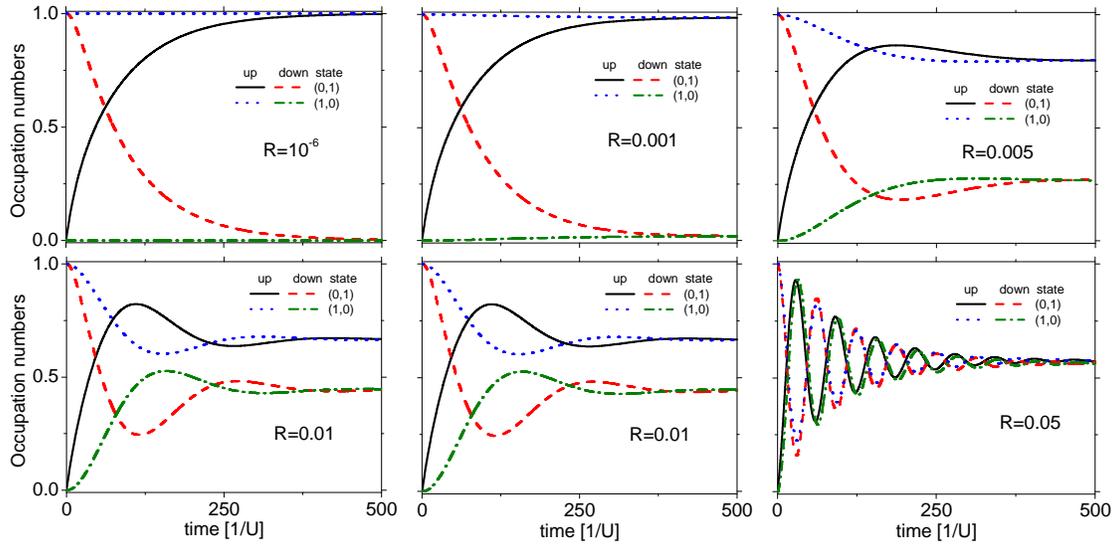}
  \caption{\label{Fig:3}
Time dependence of the occupation numbers for different initial states $(n_{\uparrow}(0),n_{\downarrow}(0))$
calculated for the indicated values of parameter $R$. Other parameters: $\epsilon_0=-0.9$, $k_BT=0.01$.}
\end{center}
\end{figure*}
%
%\end{widetext}
%
%
%\begin{widetext}
\begin{figure*}[t]
\begin{center}
  \includegraphics[width=0.82\textwidth]{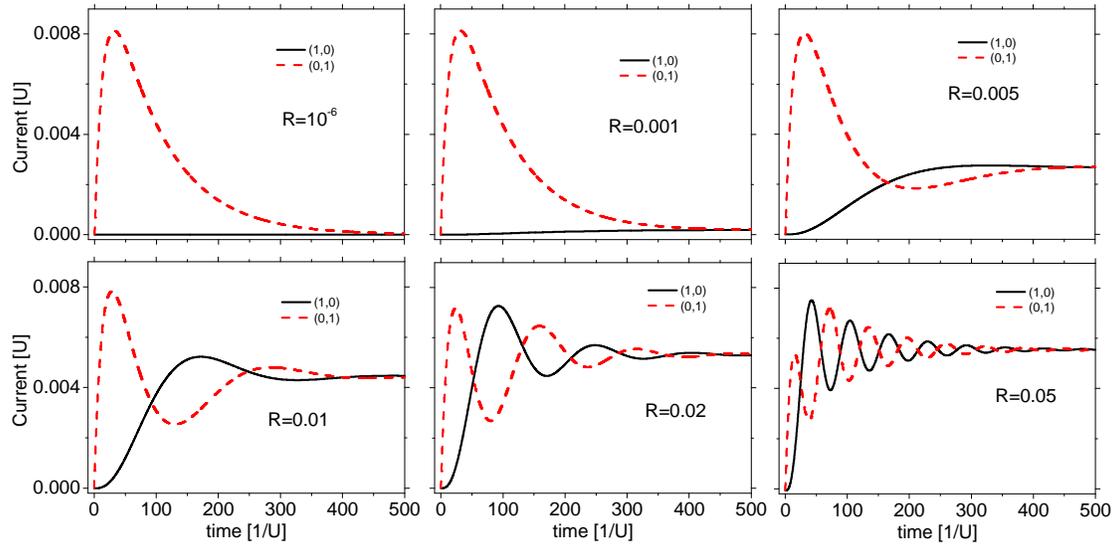}
  \caption{\label{Fig:4}
Time evolution of the charge current ($J_R$) for different initial states $(n_{\uparrow}(0),n_{\downarrow}(0))$
calculated for the indicated values of parameter $R$. Other parameters: $\epsilon_0=-0.9$, $k_BT=0.01$. }
\end{center}
\end{figure*}
%
%\end{widetext}
%
The system is then in the Coulomb blockade regime, because
$\epsilon_0<\mu_{L\sigma}, \mu_R$ and $\epsilon_0+U>\mu_{L\sigma}, \mu_R$.
In order to check how long a spin state can be maintained
before a flip to another state we have derived numerically
the rate equations (\ref{eq2}) and calculated the time dependence of
the occupation numbers $n_{\sigma}$ for given initial conditions.
In the Coulomb blockade the dot is occupied by one electron. Thus,
$\rho_{00}(0)=0=\rho_{22}(0)$, and we can write the initial dot's state as
$|n_{\uparrow}(0), n_{\downarrow}(0)\rangle=\{|1,0\rangle,|0,1\rangle\}$
or as a superposition of spin-up and spin-down states.
When the intradot spin-flip processes are neglected the state $|1,0\rangle$
is very stable, whereas the spin-down state $|0,1\rangle$ evolves into the
state $|0,1\rangle$ after some time\cite{chi}.
For a typical quantum dot at very low temperature this time can be long enough for
performing many qubit operations. However, the spin-flip effect may greatly
shorten this time scale. In Fig.\ref{Fig:2} we plot the time evolution of the
dot's occupation numbers for nonzero $R$ and for the dot's initial state $|0,1\rangle$.
One can notice that even weak spin-flip processes have a significant impact
on these quantities and hence on the spin lifetime.
We have also estimated the spin lifetime for different values of the parameter
$R$ assuming $U=10meV$, $k_BT=0.01U$ ($T\sim 1K$). The results obtained are shown in Tab.\ref{Tab:1}.
%Table
%
 \begin{table}[h]%[H] add [H] placement to break table across pages
 \caption{Calculated spin lifetime corresponding to the given value of parameter $R$\label{Tab:1}}
 \begin{ruledtabular}
 \begin{tabular}{c|c|c|c|c|c|c}
  $R$[U]    & $0$ & $10^{-6}$ & $10^{-5}$ & $10^{-3}$ & $10^{-2}$ & $10^{-1}$  \\
  $\tau$[s] & $0.6$ & $6\cdot10^{-9}$ & $6\cdot10^{-10}$ & $6\cdot10^{-12}$ & $6\cdot10^{-13}$ & $6\cdot10^{-14}$ \\
  \end{tabular}
  \end{ruledtabular}
 \end{table}
When the spin-flip strength is larger than $R\gtrsim 0.01$ ($0.1 meV$ for the assumed parameters)
the spin lifetime is too short for qubit manipulation even using (picosecond) optical methods.
The time evolution of each component of the occupation numbers
is a periodic function of the time. At the times $T=n\pi\frac{U}{R}$ ($n=1,2,...$)
the initial dot's state is restored. When the dot is initially occupied by
a spin-up electron the time evolution is analogous to that when the initial state is
$|0,1\rangle$ after making the transition $n_{\uparrow}(0,1)\rightarrow n_{\downarrow}(1,0)$
and $n_{\downarrow}(0,1)\rightarrow n_{\uparrow}(1,0)$. With increasing $R$ the time evolution
of the occupation numbers is qualitatively the same, however, the time scale is correspondingly changed (shorter).

Finally, we examine the influence of the spin-flip scattering on the read out
process. After performing manipulations on a qubit state the dot's energy level
is further lowered until $\mu_{L\uparrow}>\epsilon_0+U>\mu_R$. Now, spin-up electrons
can take part in the transport. Regardless of the spin relaxation processes, when
the dot's level is initially occupied by a spin-down electron, an electron
with spin up can tunnel through the dot (onto the dot from the left lead and further to the right lead)
and a finite current arises.
However, when a spin-up electron resides in the QD, the spin-down electron can not
tunnel onto the dot because $\epsilon_0+U>\mu_{L\downarrow}$, whereas the spin-up electron
can not leave the dot because $\epsilon_0$ is beneath the transport window.
As a result the current is suppressed. That is why measurements of the charge current
should give us information about the dot's state.
%%%%%%%%%%%%

When $R=0$ a finite current can be observed when dot was initially
occupied by a spin-down electron, whereas the current vanishes for
the state $|1,0\rangle$\cite{chi}. After a sufficiently long time the state $|0,1\rangle$
finally evolves into the state $|1,0\rangle$ and the charge current is suppressed.
However the above scenario ceases to hold when spin-flip scattering is
included. Due to the spin-flip process, now, the initial dot's state $|1,0\rangle$ ($|0,1\rangle$)
can be switched to the other spin state
and the current can be partially enhanced (suppressed) as is shown in Fig.\ref{Fig:4}. This switching mechanism leads
to oscillations in occupation numbers (see Fig.\ref{Fig:3}) as well as in the current with period inversely proportional to the strength
of the spin-flip processes. These oscillations are dumped due to the dot-lead coupling and the current saturates to a certain value.

\section{Conclusions}\label{Sec:4}
%some remarks....
In conclusion, we have studied coherent transport through a single
quantum dot subjected to spin bias in the presence of intradot spin-flip
processes. We have found that sufficiently large spin-flip strength
can prevent preparation and readout dot's
spin state by means of the present techniques. However, recent experiments
have shown that the spin lifetime can reach millisecond
or second time scales\cite{elzerman,johnson,amasha} which is enough for performing many spin
operations. That is why, along with the authors of Ref.\onlinecite{chi}, we believe that a single
spin state in the (all electrically controllable) quantum dot is a good candidate for being a qubit.

\section*{Acknowledgements}
%The author thanks Prof. J. Barna\'s for critical reading the manuscript.
This work was supported partly by funds from the Polish Ministry
of Science and Higher Education, as a research project N N202
169536 in the years 2009-2011.

\section*{Appendix: Rate Equations}
To obtain the rate equations (\ref{eq2}) we adopt the formalism presented in Ref.\cite{dong}.
Specifically, we express the dot's operator in terms of Hubbard
operators:
\begin{eqnarray}\label{eq2b}
q_{\sigma}=|0\rangle\langle\sigma|+\sigma|\bar{\sigma}\rangle\langle 2|
\end{eqnarray}
represented by four possible electron states in each
dot
which satisfy the following completeness relations:
\begin{equation}\label{eq3}
|0\rangle\langle
0|+\sum_{\sigma}\limits|\sigma\rangle\langle\sigma|+|2\rangle\langle
2|=\check{1}.
\end{equation}
Furthermore, the set of auxiliary operators is introduced and the dot operators are
expressed by means of the slave-boson and pseudofermion operators\cite{dong}.

From the
definitions of the Dirac brackets one is able to find the commutation (and
anticommutation) rules for new operators.\cite{guillou}
In the slave-boson representation the Hamiltonian of the system (\ref{eq1}) acquires the form
\begin{eqnarray}\label{eq7}
\hat{H}&=&\sum_{\mathbf k\alpha\sigma} \varepsilon_{{\mathbf
k}\alpha \sigma}c^{\dagger}_{{\mathbf k}\alpha \sigma} c_{{\mathbf
k}\alpha \sigma}+
\sum_{\sigma}\limits\epsilon_{\sigma}(f^\dag_{\sigma}f_{\sigma}
   + d^{\dag}d)
      \nonumber \\
   &+& R(f_{\uparrow}^\dag f_{\downarrow}+{\rm H.c.}) + Ud^{\dag}d
   \nonumber \\
   &+& \sum_{{\mathbf k}\alpha}\sum_{\sigma}\limits [V_{\sigma}^\alpha
   c^\dag_{{\mathbf k}\alpha\sigma}(e^{\dag}f_{\sigma}+\sigma f_{\bar{\sigma}}^{\dag}d)
   +\rm H.c.].
   \end{eqnarray}
Here, $b^{\dag}$ is the slave-boson operator which creates an
empty state in the dot, $f^{\dag}_{\sigma}$ is a peudofermion operator which
creates a singly occupied state with an electron with spin
$\sigma$, whereas $d^{\dag}$ creates a doubly occupied state with an
electron with spin $\sigma$ and another electron with spin
$\bar{\sigma}$ in the dot.
In the slave-particle representation the density matrix elements
are written in the following way:
$\hat{\rho}_{00}=e^{\dag}e$,
$\hat{\rho}_{\sigma\sigma}=f_{\sigma}^{\dag}f_{\sigma}$,
$\hat{\rho}_{22}=d^{\dag}d$. Here, the statistical expectations of
the density matrix elements ($\rho_{nn}\equiv\langle\hat{\rho}_{nn}\rangle$ with $n=0, \sigma, 2$) give the occupation
probabilities of the quantum dot being empty, singly
occupied by an electron with spin $\sigma$, and doubly occupied,
respectively.

To derive the rate equations we start from the von Neumann
equation for the density matrix operator:
\begin{eqnarray}\label{eq9}
 \dot{\hat{\rho}}=i[H,\hat{\rho}],
\end{eqnarray}
where
$\hat{\rho}=(\hat{\rho}_{00},\hat{\rho}_{\sigma\sigma},\hat{\rho}_{\sigma\bar{\sigma}},
\hat{\rho}_{22})^T$.
Furthermore, the averaged equations for density matrix elements are
expressed by means of dot-lead Green functions.
and using the Langreth theorem\cite{jauho}, the dot-lead
Green functions can be set down by means of dot's Green functions and free leads'
Green functions\cite{dong,trochaPRB10}.
The Green functions of the dot, in time space, are defined in the following way
$G_{\sigma\sigma'}(t,t')=\langle\langle
q_{\sigma}(t)|q_{\sigma'}^{\dag}(t')\rangle\rangle=\langle\langle
e^{\dag}(t)f_{\sigma}(t)|f_{\sigma'}^{\dag}(t')e(t')\rangle\rangle+\sigma\sigma'\langle\langle
f_{\bar{\sigma}}^{\dag}(t)d(t)|d^{\dag}(t')f_{\bar{\sigma'}}(t')\rangle\rangle=G_{e\sigma\sigma'}(t,t')+G_{d\sigma\sigma'}(t,t')$.
Other parts of $G_{\sigma\sigma'}(t,t')$ vanish for $t'=t$, and thus
are omitted as we are interested in the $t'=t$ case.
Now, we are able to express the rate equations by means of the dot's Green's functions and these can be written in the following form:
\begin{widetext}
\begin{eqnarray}\label{eq15}
 \dot{\rho}_{00}&=&-\frac{i}{2\pi}\int d\omega\sum_{\alpha\sigma}[\Gamma_{\sigma}^{\alpha}
 f^{\alpha}(\omega)G^>_{e\sigma\sigma}(\omega)
 +\Gamma_{\sigma}^{\alpha}
 (1-f^{\alpha}(\omega))G^<_{e\sigma\sigma}(\omega)]
\nonumber \\
\dot{\rho}_{\sigma\sigma}&=&\frac{i}{2\pi}\int
d\omega\sum_{\alpha}[\Gamma_{\sigma}^{\alpha}
 f^{\alpha}(\omega)G^>_{e\sigma\sigma}(\omega)
 +\Gamma_{\sigma}^{\alpha}
 (1-f^{\alpha}(\omega))G^<_{e\sigma\sigma}(\omega)
 -\Gamma_{\bar{\sigma}}^{\alpha}
 f^{\alpha}(\omega)G^>_{d\sigma\sigma}(\omega)
 %\nonumber \\
 -\Gamma_{\bar{\sigma}}^{\alpha}
 (1-f^{\alpha}(\omega))G^<_{d\sigma\sigma}(\omega)]
 \nonumber \\
 &+& iR(\rho_{\sigma\bar{\sigma}}-\rho_{\bar{\sigma}\sigma})
\nonumber \\
\dot{\rho}_{\sigma\bar{\sigma}}|&=&\frac{i}{4\pi}\int
d\omega\sum_{\alpha}\{(\Gamma_{\sigma}^{\alpha}+\Gamma_{\bar{\sigma}}^{\alpha})[
 f^{\alpha}(\omega)G^>_{e\sigma\bar{\sigma}}(\omega)
 +
 (1-f^{\alpha}(\omega))G^<_{e\sigma\bar{\sigma}}(\omega)
 -
 f^{\alpha}(\omega)G^>_{d\bar{\sigma}\sigma}(\omega)
 %\nonumber \\
 -
 (1-f^{\alpha}(\omega))G^<_{d\bar{\sigma}\sigma}(\omega)]\}
 \nonumber \\
 &+& iR(\rho_{\sigma\sigma}-\rho_{\bar{\sigma}\bar{\sigma}})
\nonumber \\
\dot{\rho}_{22}&=&\frac{i}{2\pi}\int
d\omega\sum_{\alpha\sigma}[\Gamma_{\sigma}^{\alpha}
f^{\alpha}(\omega)G^>_{d\bar{\sigma}\bar{\sigma}}(\omega)
+\Gamma_{\sigma}^{\alpha}
(1-f^{\alpha}(\omega))G^<_{d\bar{\sigma}\bar{\sigma}}(\omega)]
\end{eqnarray}
\end{widetext}

Now, we only need to know the Green's function of the dot. This is derived in the weak coupling
approximation from the corresponding equation of motion
for the dot's operators. Thus, we obtained
\begin{widetext}
\begin{eqnarray}\label{eq16}
G^<_{e\sigma\sigma}(\omega)&=&i\pi\rho_{\sigma\sigma}\{\delta[\omega-(\epsilon_0+R)]+\delta[\omega-(\epsilon_0-R)]\}
%\nonumber \\
+ i\pi\rho_{\bar{\sigma}\sigma}\{\delta[\omega-(\epsilon_0+R)]-\delta[\omega-(\epsilon_0-R)]\},
\nonumber \\
G^>_{e\sigma\sigma}(\omega)&=&-i\pi\rho_{00}\{\delta[\omega-(\epsilon_0+R)]+\delta[\omega-(\epsilon_0-R)]\},
\nonumber \\
G^<_{d\bar{\sigma}\bar{\sigma}}(\omega)&=&i\pi\rho_{22}\{\delta[\omega-(\epsilon_0+U+R)]+\delta[\omega-(\epsilon_0+U-R)]\},
\nonumber \\
G^>_{d\bar{\sigma}\bar{\sigma}}(\omega)&=&-i\pi\rho_{\bar{\sigma}\bar{\sigma}}\{\delta[\omega-(\epsilon_0+U+R)]+\delta[\omega-(\epsilon_0+U-R)]\}
%\nonumber \\
+ i\pi\rho_{\sigma\bar{\sigma}}\{\delta[\omega-(\epsilon_0+U+R)]-\delta[\omega-(\epsilon_0+U-R)]\},
\nonumber \\
G^<_{e\sigma\bar{\sigma}}(\omega)&=&i\pi\rho_{\sigma\bar{\sigma}}\{\delta[\omega-(\epsilon_0+R)]+\delta[\omega-(\epsilon_0-R)]\}
%\nonumber \\
+i\pi\rho_{\bar{\sigma}\bar{\sigma}}\{\delta[\omega-(\epsilon_0+R)]-\delta[\omega-(\epsilon_0-R)]\},
\nonumber \\
G^>_{e\sigma\bar{\sigma}}(\omega)&=&-\pi\rho_{00}\{\delta[\omega-(\epsilon_0+R)]-\delta[\omega-(\epsilon_0-R)]\},
\nonumber \\
G^<_{d\bar{\sigma}\sigma}(\omega)&=&i\pi\rho_{22}\{\delta[\omega-(\epsilon_0+U+R)]-\delta[\omega-(\epsilon_0+U-R)]\},
\nonumber \\
G^>_{d\bar{\sigma}\sigma}(\omega)&=&i\pi\rho_{\sigma\bar{\sigma}}
\{\delta[\omega-(\epsilon_0+U+R)]+\delta[\omega-(\epsilon_0+U-R)]\}
%\nonumber \\
-i\pi\rho_{\sigma\sigma}\{\delta[\omega-(\epsilon_0+U+R)]-\delta[\omega-(\epsilon_0+U-R)]\}.
\nonumber \\
\end{eqnarray}
\end{widetext}
To derive these Green functions we assumed no coupling to the leads ($V_{\sigma}^\alpha=0$)
and that the leads are taken to be in local thermal equilibrium. In contrast to Refs.\cite{rudzinski1,dong,glazman} we
did not assume vanishing $R$ during the calculation of the above Green's functions. Thus, our equations are not limited to small
values of the spin-flip strengths ($R\ll\Gamma$). Moreover, the form of these Green's functions clearly shows that spin-flip processes
lead to splitting of the dot's energy level.
Finally, connecting
Eqs.(\ref{eq16}) with Eqs.(\ref{eq15}) we arrive at the coupled set of differential equations (\ref{eq2}).


\begin{thebibliography}{25}
%1

\bibitem{loss1} D. Loss, and D. P. DiVincenzo, Phys. Rev. A {\bf 57} (1998) 120.


\bibitem{koppens} F. H. L. Koppens, C. Buizerk, K. J. Tielrooij, I. T. Vink, K. C.
Nowack, T. Meunier, L. P. Kouwenhoven, and L. M. K. Vandersypen, Nature (London) {\bf 442} (2006) 766.

%3
\bibitem{press} D. Press, T. D. Ladd, B. Y. Zhang, and Y. Yamamoto, Nature
(London) {\bf 456} (2008) 218.

%4
\bibitem{xu} X. D. Xu, Y. W. Wu, B. Sun, Q. Huang, J. Chen, D. G. Steel, A.
S. Bracker, D. Gammon, C. Emary, and L. J. Sham, Phys. Rev.
Lett. {\bf 99} (2007) 097401.


\bibitem{nowack} K. C. Nowack, F. H. L. Koppens, Yu. V. Nazarov, and L. M. K.
Vandersypen, Science {\bf 318} (2007) 1430.

\bibitem{elzerman} J. M. Elzerman, R. Hanson, L. H. Willems van Beveren, B.
Witkamp, L. M. K. Vandersypen, and L. P. Kouwenhoven, Nature
(London) {\bf 430} (2004) 431.

\bibitem{hanson} R. Hanson, L. H. van Beveren Willems  , I. T. Vink, J. M. Elzerman,
W. J. M. Naber, F. H. L. Koppens, L. P. Kouwenhoven, and
L. M. K. Vandersypen, Phys. Rev. Lett. {\bf 94} (2005) 196802.

%6
\bibitem{chi} F. Chi and Q. -F. Sun, Phys. Rev. B {\bf 81} (2010) 075310.


\bibitem{frolov1} S. M. Frolov, A. Venkatesan, W. Yu, and J. A. Folk, and W. Wegscheider, Phys. Rev. Lett. {\bf 102} (2009) 11602.


\bibitem{frolov2} S. M. Frolov, S. L\"uscher, W. Yu, Y. Ren, J. A. Folk, and W. Wegscheider, Nature {\bf 458} (2009) 868.


\bibitem{lu} H. Z. Lu and S. -Q. Shen, Phys. Rev. B {\bf 77} (2008) 235309.


\bibitem{fujisawa} T. Hayashi, T. Fujisawa, H. D. Cheong, Y. H. Jeong, and
Y. Hirayama, Phys. Rev. Lett. {\bf 91} (2003) 226804.

\bibitem{taylor} J. M. Taylor, J. R. Petta, A. C. Johnson, A. Yacoby, C. M. Marcus,
and M. D. Lukin, Phys. Rev. B {\bf 76} (2007) 035315.

\bibitem{rudzinski1} W. Rudzi\'nski and J. Barna\'s, Phys. Rev. B {\bf 64} (2001) 085318.

\bibitem{rudzinski2} W. Rudzi\'nski, J. Phys.: Condens. Matter {\bf 21} (2009) 046005.

\bibitem{nazarov} A. V. Khaetskii and Y. V.  Nazarov, Phys. Rev. B {\bf 61} (2000) 12639.

\bibitem{johnson} A. C. Johnson, J. R. Petta, J. M. Taylor, A. Yacoby, M. D. Lukin, C. M. Marcus, M. P. Hanson, and A. C. Gossard, Nature {\bf 435} (2005) 925.

\bibitem{amasha} S. Amasha, K. MacLean, I. P. Radu, D. M. Zumbühl, M. A.
Kastner, M. P. Hanson, and A. C. Gossard, Phys. Rev. Lett. {\bf 100} (2008) 046803.

%SRtimes


\bibitem{tarucha} R. Hanson, L. P. Kouwenhoven, J. R. Petta, S. Tarucha, and L.
M. K. Vandersypen, Rev. Mod. Phys. {\bf 79} (2007) 1217.

\bibitem{meunier} T. Meunier, I. T. Vink, L. H. Willems van Beveren, K. -J. Tielrooij,
R. Hanson R, F. H. L. Koppens, H. P. Tranitz, W. Wegscheider,
L. P. Kouwenhoven, and L. M. K. Vandersypen, Phys. Rev. Lett. {\bf 98} (2007) 126601.

\bibitem{heiss} D. Heiss, S. Schaeck, H. Huebl, M. Bichler, G. Abstreiter,
J. J. Finley, D. V. Bulaev, and D. Loss, Phys. Rev. B
{\bf 76} (2007) 241306(R).


\bibitem{weymann} I. Weymann and J. Barna\'s, Phys. Rev. B {\bf 73} (2006) 205309.




%Spin-Bias

\bibitem{hirsch} J.E. Hirsch, Phys. Rev. Lett. {\bf 83} (1999) 1834.

\bibitem{wang2} D.-K. Wang, Q.-F. Sun, and H. Guo, Phys. Rev. B {\bf 69} (2004) 205312.


%spin-orbit
\bibitem{nazarov} A. V. Khaetskii and Y. V. Nazarov, Phys. Rev. B {\bf 61} (1999) 12639(R);
C. F. Destefani, S. E. Ulloa, and G. E. Marques, Phys. Rev. B {\bf 69} (2004) 12502;
E. Tsitsishvili, G. S. Lozano, A. O. Gogolin, Phys. Rev. B {\bf 70} (2004) 115316.

\bibitem{loss2} H.-A. Engel and D. Loss, Phys. Rev. Lett. {\bf 86} (2000) 4648.

%RateEquations
\bibitem{dong} B. Dong, H. L. Cui, and X. L. Lei, Phys. Rev. B {\bf 69} (2004) 035324.


\bibitem{guillou} J.C. Le Guillou, E. Ragoucy, Phys. Rev. B {\bf 52}, 2403 (1995).

\bibitem{jauho} H. Haug, A.-P. Jauho, \emph{Quantum Kinetics in Transport and Optics of Semiconductors},
Springer Berlin Heidelberg New York, Second Edition (2008).

\bibitem{trochaPRB10} P. Trocha, Phys. Rev. B {\bf 82} (2010) 115320.

\bibitem{glazman} L. I. Glazman and K. A. Matveev, Pis'ma Zh. Eksp. Teor. Fiz.
{\bf 48} (1998) 403; JETP Lett. {\bf 48} (1998) 445.



\end{thebibliography}
\end{document}